\documentclass{revtex4}
\usepackage{amsmath}
\usepackage{amssymb}
\usepackage{graphicx,color}

\begin{document}

\title{Nonperturbative quantization technique for QCD}

\author{Vladimir Dzhunushaliev}
\email{vdzhunus@krsu.edu.kg} 
\affiliation{
Institute for Basic Research, 
Eurasian National University, 
Astana, 010008, Kazakhstan
Institut f\"ur Physik, 
Universit\"at Oldenburg, Postfach 2503, 
D-26111, Oldenburg, Germany}

\begin{abstract}
Heisenberg nonperturbative quantization technique for quantum chromodynamics is applied. In such approach the nonperturbative quantization is based on Yang - Mills equations applied for the quantum field operator $\hat A^B_\mu$. It is shown that such equation is equivalent to an infinite equations set for all Green functions. Various approximate methods for solving the infinite equations set are discussed. 
\end{abstract}


\maketitle

\section{Introduction}

There is a widespread opinion that nonperturbative quantization is absolutely unknown procedure in physics. We think that it is not so. In 50th Heisenberg has offered the method for the nonperturbative quantization of a nonlinear spinor field \cite{heisenberg}. Following to Heisenberg nonperturbative quantized operators of a quantum field can be calculated using corresponding field equation(s) for this theory. \textcolor{blue}{\emph{The main idea is that the  field equation(s) are written for quantum field operators.}} The problem is that nobody knows how one can solve such equation(s). Fortunately the equation(s) is equivalent to an infinite differential equations set for Green functions. The problem with solving such infinite differential equations set is very difficult also, nevertheless it is easier than the problem arising with differential equations for operators.

One of the objections against such nonperturbative quantization procedure is that Heisenberg has applied it for the quantization of a nonlinear Dirac equation with selfinteraction like $\left| \psi \right|^4$. Some physicists say that such quantization is senseless because this nonlinear theory is nonrenormalizable. But such comment is not true because the notion of renormalization can be applied for perturbative quantized theories only. It means that the theory can be nonrenormalizable but it can be quantized, i.e. the nonrenormalizability of the theory does not mean that the theory can not be quantized. 

\section{Heisenberg nonperturbative quantization technique for QCD}

According Heisenberg \cite{heisenberg} quantum field operators $\hat A^B_\mu$ of the SU(3) gauge theory obey to operator Yang - Mills equations 
\begin{equation}
	D_\nu \hat F^{B \mu \nu} = 0 
\label{1-10}
\end{equation}
here for the simplicity we consider a pure Yang - Mills theory without sources; 
$\hat F^B_{\mu \nu} = \partial_\mu \hat A^B_\nu - \partial_\nu \hat A^B_\mu + 
g f_{BCD} \hat A^C_\mu \hat A^D_\nu$ is the operator of field strength; $\hat A^B_\mu$ is the operator of SU(3) gauge potential; $B,C,D = 1, \ldots ,8$ are the SU(3) color indices; $g$ is the coupling constant; $f^{BCD}$ are the structure constants for the SU(3) gauge group; $D_\nu$ is the gauge derivative.  \textcolor{blue}{\emph{The nonperturbative quantization for QCD means that the quantum field operators $\hat A^B_\mu$ are defined by operator Yang - Mills equations \eqref{1-10}.}}

How we can solve this equation ? The answer is that the operator equation \eqref{1-10} is equivalent to an infinite equations set 
\begin{eqnarray}
	\left\langle Q \left| D_\nu \hat F^{B \mu \nu}(x) 
	\right| Q \right\rangle &=& 0 ,
\label{1-20}\\
	\left\langle Q \left| \hat A^{C_1}_{\rho_1}(x_1) 
	D_\nu \hat F^{B \mu \nu}(x) 
	\right| Q \right\rangle &=& 0 ,
\label{1-30}\\
	\cdots &=& 0 ,
\label{1-40}\\
	\left\langle Q \left| \hat A^{C_1}_{\rho_1}(x_1) \cdots 
	\hat A^{C_n}_{\rho_n}(x_n) D_\nu \hat F^{B \mu \nu}(x)) 
	\right| Q \right\rangle &=& 0 ,
\label{1-50}\\
	\cdots &=& 0 
\label{1-60}
\end{eqnarray}
where $\left. \left|Q \right. \right\rangle$ is some quantum state, for example, it can be a quantum state for a glueball, flux tube, proton and so on. Schematically the first equation \eqref{1-20} has 
$\left\langle Q \left| A \right| Q \right\rangle$ and 
$\left\langle Q \left| A^3 \right| Q \right\rangle$ terms; the second equation
\eqref{1-30} has $\left\langle Q \left| A^2 Q \right| \right\rangle$ and 
$\left\langle Q \left| A^4 \right| Q \right\rangle$ terms; the n-th equation  
\eqref{1-50} has $\left\langle Q \left| A^n \right| Q \right\rangle$ and   
$\left\langle Q \left| A^{n+2} \right| Q \right\rangle$ terms and so on. Thus all equations are linked up and this is the main problem to solve such infinite equations set. Similar equations set one can find in statistical physics and turbulence. On the perturbative language Eq's \eqref{1-20}-\eqref{1-60} are Dyson - Schwinger equations but usually they are written on the Feynman diagrams language. 

\emph{Comment. In Eq's \eqref{1-20}-\eqref{1-60} there are products like 
$\hat A^n(x)$. For the perturbative approach such product leads to singularities because the product of field operators in one point is poorly defined. We have to point out on the difference between such products by perturbative and nonperturbative quantization: for the nonperturbative quantization such singularities may be much softer or be absent in general \cite{heisenberg}.}

The matter is that in the perturbative approach we have moving particles (quanta). The communication between two points is carried by such quanta. Consequently the correlation between these points (Green function) is not zero if only the exchange with quanta is possible. It means that corresponding Green function is not zero inside of a light cone and is zero outside of them. Such kind of functions have to be singular on the light cone (it is well known). But for the nonperturbative case it is not the case: for the self-interacting nonlinear fields may exist static configurations. In this case the correlation (Green function) is not zero outside of the light cone. This is the reason why the Green function for the nonperturbative case is not so singular in the comparison with the perturbative case. 

With high probability such equations set can not be solved analytically. Only one way does exist: to tear the infinite equations set by using some physically meaningful assumptions. For example, one can assume that 
\begin{equation}
	\left\langle Q \left| A^n \right| Q \right\rangle 
	\approx \left\langle Q \left| A^{n-2} \right| Q \right\rangle
	\left( 
		\left\langle Q \left| A^2 \right| Q \right\rangle - C_2 
	\right) + \cdots 
\label{1-70}
\end{equation}
or something like that; here $C_2$ is some constant and the physical
consequences of $C_2 \neq 0$ will be discussed in section \ref{possible}. After
that we will have a finite equation set which can be solved analytically or
(that is more probable) numerically. Such procedure is often used in the theory
of turbulent fluid and statistical physics. 

Concerning to the equivalence of the operator equation \eqref{1-10} and infinite equations set \eqref{1-20}-\eqref{1-60} one can say that the set \eqref{1-20}-\eqref{1-60} determines both operators $\hat A^B_\mu$ and the quantum state $\left. \left|Q \right. \right\rangle$. 

\section{Possible ways for solving Eq's \eqref{1-20}-\eqref{1-60}}
\label{possible}

Generally speaking similar equations set are well known in statistical and turbulent physics. One can use these methods for solving such equations in QCD. 

One possible way is mentioned above. In such approach we decompose $n-$th Green function 
\begin{equation}
	G_n = G^{B_1, B_2, \cdots , B_n}_{\mu_1, \mu_2, \cdots , \mu_n} 
	(x_1, x_2 \cdots , x_n) = 
	\left\langle Q \left| 
		A^{B_1}_{\mu_1}(x_1) A^{B_2}_{\mu_2}(x_2) \cdots A^{B_n}_{\mu_n}(x_n) 
	\right| Q \rangle\right| 
\label{2-10}
\end{equation}
on the linear combination of the products 
\begin{equation}
\begin{split}
	G_n(x_1, x_2 \cdots , x_n) \approx & G_{n-2}(x_3, x_4 \cdots , x_n)
	\textcolor{blue}{\left[ G_2(x_1, x_2) -C_2 \right]} + 
\\
	&
	\left( \text{permutations of $x_1,x_2$ with
	$x_3, \cdots , x_n$} \right) + 
\\
	&	
	G_{n-3} G_3 + \cdots 
\end{split}
\label{2-20}
\end{equation}
In such a way one can cut off the infinite equation set
\eqref{1-20}-\eqref{1-60}. The term $G_2(x_1, x_2) -C_2$ may lead
to very interesting physical consequences: the Green function $G_n$ will be
zero for \textcolor{blue}{\emph{nonzero}} $G_2 = C_2$. It means that now a
vacuum state is realized not for all zero Green functions $G_i = 0$ but
the vacuum state is realized by some nonzero Green functions, for example, 
by $G_2\neq 0$ (vacuum displacement).

One can also try to solve approximately these equations. For example, one can choose some functional (action, for example) and average it. After that one can simplify 2-point Green function using some assumptions. For example, one can assume that $G_2(x_1, x_2)$ can be approximately decomposed on the product of two scalar functions 
\begin{equation}
	G^{AB}_{\mu \nu}(x_1, x_2) \approx C^{AB}_{\mu \nu} \phi^*(x_1) \phi(x_2)
\label{2-30}
\end{equation}
where $C^{AB}_{\mu \nu}$ is a numerical factor. Thereafter one can assume that 4-point Green function in one point $x$ is a bilinear combination of 2-points Green functions. Schematically it is 
\begin{equation}
	G_4(x,x,x,x) \approx G_2^2(x,x) - C_2 G_2^2(x,x) + B
\label{2-35}
\end{equation}
where $C_2,B$ are some constants. In such scalar approximation we take away the
color $(A,B)$ and Lorentzian $(\mu, \nu)$ indexes. As the consequence we reduce
initial number of degrees of freedom to only one - scalar field. As the result
we obtain an effective Lagrangian for the scalar field. Varying with the scalar
field we obtain Euler - Lagrange equations for such approximate approach. 

\emph{Comment. In such simplification the scalar field $\phi$ appears after the quantization. It means that we should not quantize such kind of scalar fields. 
}

Also one can suppose that in some physical phenomenon the physical degrees of freedom split on two kinds of degrees of freedom. The first ones are almost classical degrees of freedom but the second ones are quantum ones. It can occur in a flux tube (stretched between quark and antiquark): a longitudinal color electric field (created by quark-antiquark) is in a classical phase and quantum degrees of freedom (basically they are color magnetic field, according to dual QCD picture) confine the classical color electric field into a tube. 

\section{Discussion}

There are some reasons why in the course of long time the Heisenberg approach undeservedly has been forgotten: 
\begin{itemize}
	\item About at the same time well working a perturbative quantization technique has appeared - Feynman diagrams. 
	\item The calculations with Feynman diagrams were much easier and yielded fine results in agreement with the experiments in contrast with very complicated calculations in Heisenberg approach. 
	\item At the perturbative quantization the nonlinear spinor theory of Dirac  is not renormalized. It became an obstacle for the further consideration of the nonlinear spinor theory of Dirac for the physicists working with the perturbative approach. Though we should understand that the nonrenormalizability does not mean the nonquantizability.
\end{itemize}
It is necessary to note that  many problems exist in this nonperturbative way. The main problem from the author point of view is to determine what kind of algebra of quantum field operators follows from the operator Yang - Mills equations ? The matter is that for the perturbative approach such algebra is defined through canonical commutation relationships for free quantum field operators. \textcolor{blue}{\emph{But for interacting quantum field operators these canonical commutation relationships are not correct !}} The question is: how we can determine the algebra for \textcolor{blue}{\emph{interacting}} quantum field operators ? We think that it should be follow from the operator Yang - Mills equations \eqref{1-10}. 

Other problems are: 
\begin{itemize}
	\item mathematical equivalence between the operator Yang - Mills equations and the infinite equations set for Green functions;
	\item the determination of quantum states from the infinite equations set for Green functions;
	\item the proof of convergence of the solution. 
\end{itemize}
In conclusion we would like to turn our attention to the difference between quantum fields in perturbative and nonperturbative approaches. In the first perturbative case any quantum field is constructed from quanta (their elementary excitations). But in the second nonperturbative case the quantum field \emph{can not be presented as a cloud of quanta.} The situation here is more similar to a turbulent fluid when we move together with it in such a way that it is visible only a rest fluctuating fluid. 

\section*{Acknowledgements}

I am grateful to the Research Group Linkage Programme of the Alexander  von
Humboldt Foundation for the support of this research. Special thanks to J. Kunz for invitation to Universit\"at Oldenburg for research.


\begin{thebibliography}{99}

\bibitem{heisenberg}
W. Heisenberg, \textit{Introduction to the unified field theory of
elementary particles.}, Max - Planck - Institut f\"ur Physik und
Astrophysik, Interscience Publishers London, New York, Sydney,
1966. 


\end{thebibliography}
\end{document}